\documentclass[reprint,amsmath,amssymb,aps,showpacs]{revtex4-1}  
  
\usepackage{graphicx}
\usepackage{dcolumn}
\usepackage{bm}  
  
\begin{document}

\title{
Finite size scaling analysis of a 
nonequilibrium phase transition \\ in the naming game model 
} 

\author{ E. Brigatti and A. Hern\'andez}

\affiliation{Instituto de F\'{\i}sica, Universidade Federal do Rio de Janeiro, 
Av. Athos da Silveira Ramos, 149,
Cidade Universitária, 21941-972, Rio de Janeiro, RJ, Brazil}
\address{e-mail address: edgardo@if.ufrj.br}

\begin{abstract}

We realize an extensive numerical study of the Naming Game
model with a noise term which
accounts for perturbations. 
This model displays a non-equilibrium phase transition between
an absorbing ordered consensus state, which occurs for small noise,
and a disordered phase with fragmented clusters characterized by
heterogeneous memories,
which emerges at strong noise levels. 
The nature of the phase transition is studied by means of a
 finite-size scaling analysis of the moments. 
We observe a scaling behavior typical of a discontinuous transition and 
we are able to estimate the thermodynamic limit.
The scaling behavior of the clusters size seems also compatible with this kind of transition.

\end{abstract}
 

\date{\today}
\maketitle

\section{Introduction}

The contributions of statistical physicists to 
the understanding of spreading 
of  cultural traits, opinions or conventions
of different nature are nowadays well established
\cite{loreto_review}.
Several works 
focus on the general mechanisms responsible for the ordering 
dynamics that generates  global consensus 
as an emergent phenomenon \cite{ordering}.
The introduction of minimal models
is a standard practice of the area, 
which is intended to uncover
the possible existence of some 
universal character shared by different real systems.
Such idea of universal properties is in fact supported by
the classical theory of phase transitions, which is expected 
to give at least a partial legacy 
for these non-equilibrium systems.

In this work we focus our attention to a model of conventions
spreading successfully used for describing linguistic dynamics \cite{puglisi08}.
 The model, usually called Naming Game \cite{baronchelli06}, is characterized 
 by a collective dynamics which implement a memory-based negotiation strategy, 
 where a sequence of trials shapes and reshapes the system memories, allowing for intermediate individual states and feedback effects.
These rules appear to be more realistic  than simple imitation or local majority mechanisms, commonly implemented by the use of Ising-like dynamics.
The introduction 
of a noise term, which can 
account for external or internal perturbations or agents' irresolute attitude,
generates  global consensus for small noise levels \cite{baronchelli06}.
On the other hand, strong noise conducts the system to a stationary state
characterized by several coexisting conventions.
On the basis of some analytical considerations, supported by numerical 
simulations, it was suggested that the onset of the consensus state can be 
described as a non-equilibrium phase transition giving rise to
order \cite{baronchelli07}.

The purpose of this work is to study and clearly characterize
the nature of this transition 
throughout an accurate finite-size scaling analysis. 
For this reason, we consider the implementation of the model on a
2D square lattice, where our Monte Carlo simulations are performed. 

In section 2 we will describe the details of the model
and we will introduce a comparison with prior results in the literature,
in section 3 we will report the numerical analysis for the characterization of
the phase transition and we will discuss our results.

\section{The model}

The dynamics rules defining the model are quite simple.
Each player is characterized by an inventory 
which can contain an infinite number of conventions.
In fact, it is structured by an array of potentially infinite 
cells where each cell is set on one of an infinite 
number of possible numerable states.
In the initial state players start with an empty inventory.
At each time step, a pair of agents is randomly selected.
The first agent 
selects one of its conventions 
or, if its inventory is empty, it creates a new one. 
After that, the convention 
is transmitted to the second agent.
If this last agent possesses the transmitted convention,
 with a probability $\beta$, the two agents update their inventories so as to keep only the convention 
involved in the interaction. Conversely, with probability $1-\beta$, no actions are performed by the couple of agents.
Otherwise, if the second agent does not possess the transmitted convention,
the interaction is a failure and  it adds the new convention 
in its inventory.

The game is simulated on a regular 2D square lattice with $L\times L$ sites
and periodic boundary conditions. This implementation defines a short-range interaction system, where agents communicate only with their four nearest neighbors.
The special case where $\beta=1$ corresponds to the original Naming Game 
embedded on a low-dimensional lattice. This model 
was extensively studied and it shown different convergence behaviors from
the mean-field case. 
In fact, consensus is reached by means of a coarsening process which needs less
agents' memory effort but longer convergence times than the mean-field model
(the upper critical dimension is 4) \cite{baronchelli06b}.

 The model with general values of $\beta$, which effectively generates
  the described transition, was studied only in one previous
  work \cite{baronchelli07}.
That paper studied a special case of the mean-field model
by means of an analytical approximation where agents can store a
maximum of only two different conventions. 
In this specific approximation of the original model,
it was shown that a shift from a consensus state to a 
polarized one exists at $\beta_c=1/3$. 
Simulations of the original model, i.e. with an unlimited number of 
conventions, suggested that the transition
happen at the same value of $\beta$ obtained
for the analytical approximation \cite{baronchelli07}.
This fact was inferred looking at the divergence of the
convergence time near those values.
In details, the convergence time required by the system to
reach the consensus state ($t_{conv}$),
present two different behaviors \cite{baronchelli07}: 
one for the simplified case with just two conventions,
where  $t_{conv}\propto (\beta-\beta_c)^{-1}$
and one for the case of the original model,
with an unlimited number of conventions,
where $t_{conv}\propto (\beta-\beta_c)^{-0.3}$.
Finally, using similar arguments, the authors shown 
that homogeneous random networks and
heterogeneous topologies with power-law 
degree distributions
present the same $\beta$ value for the transition
if the pair selection criterion consists in randomly
choosing a link \cite{baronchelli07}.


\section{Results and discussion}

In the following we develop a finite size scaling analysis
to clearly characterize the phase transition of the model 
implemented on a 2D square lattice.
The system presents very slow relaxation time close to the transition.
For this reason, we analyze the final state reached after
running $4 \times 10^{10}$  Monte Carlo steps.
Such very long simulations force us to adopt $L$ values limited
between 20 and 60.

\begin{figure}[h]
\begin{center}
\vspace{0.6cm}
\includegraphics[width=0.5\textwidth, angle=0]{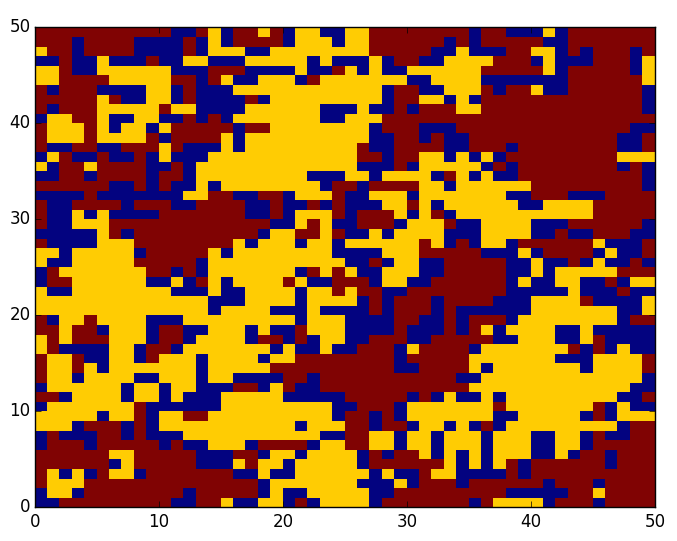}
\end{center}
\caption{\small An active and fragmented steady state ($\beta=0.32$ and $L=50$).
Two conventions 
are exchanged: in the red and yellow sites only one convention 
is present, in the blue sites the two conventions 
coexist. 
}
\label{fig_fragments}
\end{figure}

The phase transition is marked by the 
passage from an active stationary state of disordered and fragmented clusters 
to an absorbing state of a single cluster represented by the same word.
In fact, for small $\beta$ convergence
is not attained and small clusters, with one or more
different words, characterize the system.
In particular, for $\beta$ values slightly smaller than 
the one which generates a single cluster ($\beta_{c}$),
the fragmented state is characterized by the presence of just
two conventions,
in accordance with the mean-field behavior of the 
model \cite{baronchelli07} (see Figure \ref{fig_fragments}). 

\begin{figure}[h]
\begin{center}
\vspace{0.6cm}
\includegraphics[width=0.5\textwidth, angle=0]{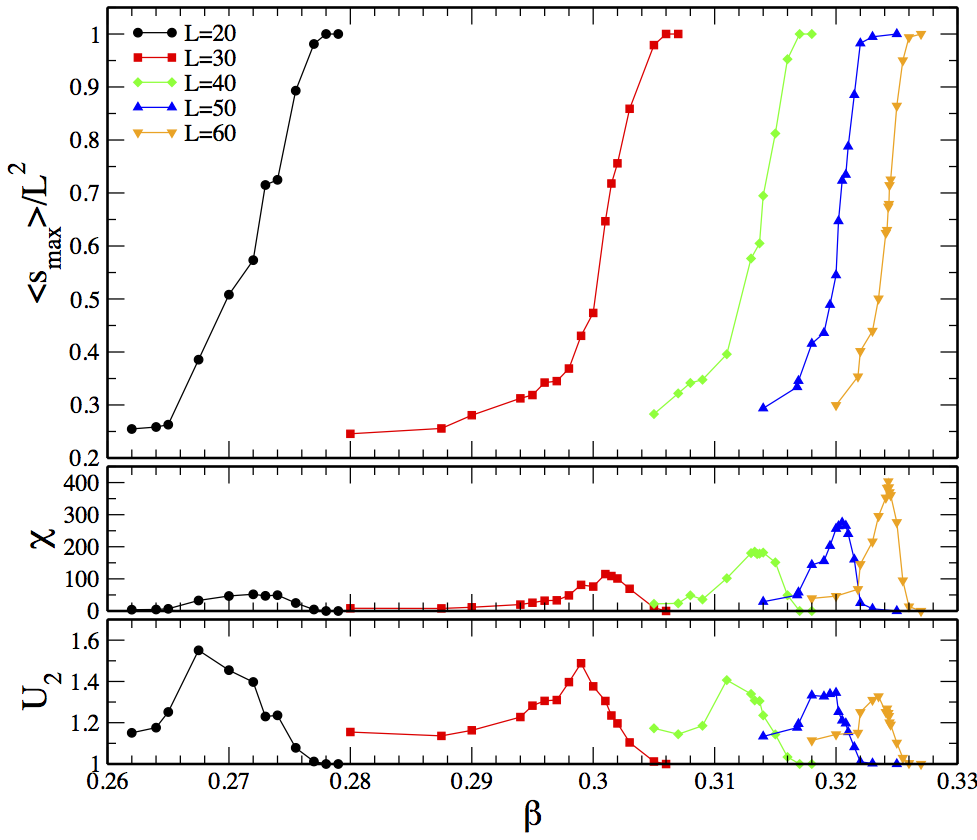}
\end{center}
\caption{\small From top to bottom: Mean of the normalized largest cluster
size, 
its variance $\chi$ and the moment ratio $U_2$
as a function of $\beta$ for different system sizes. 
Each point is averaged over $100$ simulations.}
\label{fig_phtrans}
\end{figure}

For these reasons the relative size of the largest cluster
present in the system is an ideal parameter to characterize
the transition 
\cite{castellano00,brigatti15}.
This is defined as the size of the largest cluster 
composed by agents sharing the same unique convention 
normalized over the system size: $s_{max}/L^{2}$.
In Figure \ref{fig_phtrans} we can observe its behavior 
for different values of $L$.
In accordance with typical phase transitions, 
we can observe a clear scaling looking at the rise of the critical value of $\beta_{c}(L)$ and at the characteristic steeper transitions
for larger system sizes.

For  a clear characterization of the type of the transition (continuous or discontinuous) 
and a  quantitative estimation of $\beta_c$ it is useful to evaluate the fluctuations of the size of the largest cluster:
\begin{eqnarray}
\chi=L^{2}(<s_{max}^{2}> - <s_{max}>^{2}) 
\nonumber
\end{eqnarray}
and its moment ratio (reduced cumulant) \cite{dickman98}:
\begin{eqnarray}
U_2=\frac{<s_{max}^{2}>}{<s_{max}>^{2}};
\nonumber
\end{eqnarray}
where $<\;\;>$ stands for averages over different simulations. 

As can be seen in Figure \ref{fig_phtrans}, these quantities peak  around 
$\beta_c(L)$. In particular, the maxima of the fluctuations are
characterized by  higher values for increasing values of $L$. 
In fact, the use of the maxima of these quantities has recently demonstrated to be a very robust approach for performing a finite size scaling analysis of discontinuous phase transition into absorbing states \cite{oliveira2015}.
In this case, the asymptotic transition
point (asymptotic coexistence point) can be obtained looking at the convergence of the finite size transition points $\beta_c(L)$ as estimated by the localization 
of the maxima of the fluctuations or the maxima of the moment ratio.
In both cases, the convergence  is expected to follow an algebraic behavior:
$  \beta_c(L)=\beta_c+aL^{-2}$,
which is the usual equilibrium scaling \cite{oliveira2015,privman}.

\begin{figure}[t]
\begin{center}
\includegraphics[width=0.5\textwidth, angle=0]{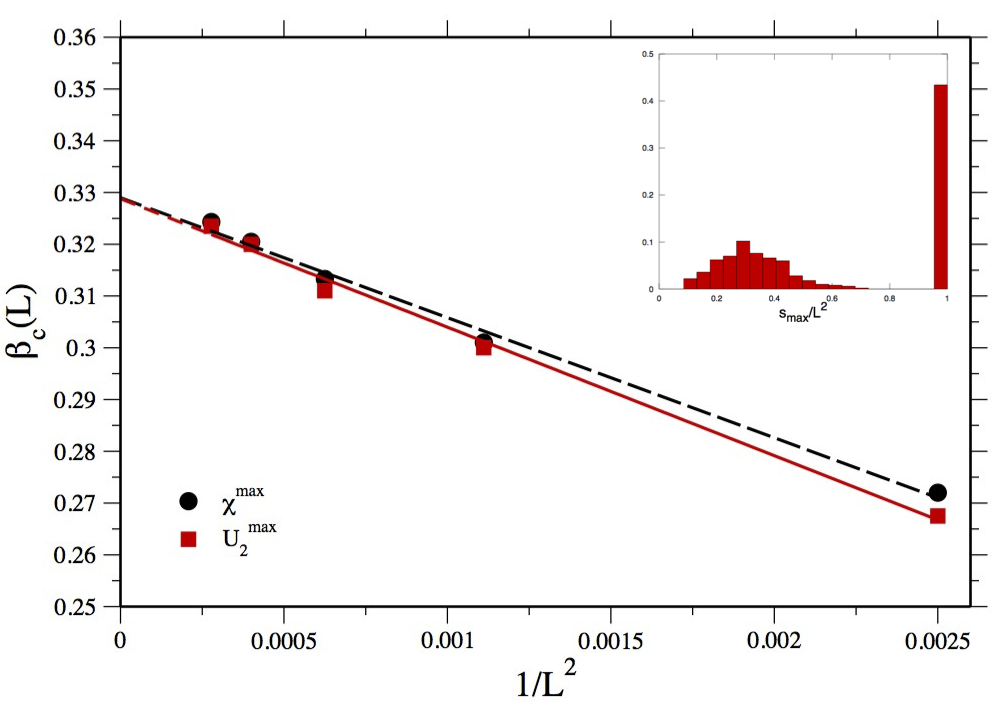}
\end{center}
\caption{\small 
Convergence to the asymptotic transition point $\beta_c$ 
of the finite-size transition points $\beta_c(L)$ measured from
the variance and the $U_2$ maxima.
In the inset, a typical example of the shape of the probability distribution function of the 
normalized largest cluster size 
near $\beta_{c}(L)$ (in this case $\beta=0.301, L=30$).
}
\label{fig_critic}
\end{figure}

These scaling laws are well verified by our data.
Looking at Figure \ref{fig_critic} we can see that the maxima positions
for $\chi$ and $U_2$ effectively decrease as $1/L^2$. An extrapolation for $L\to \infty$ yields $\beta_c= 0.329\pm0.001$  
for the two cases, an excellent agreement between them. 

\begin{figure}[h]
\begin{center}
\vspace{0.6cm}
\includegraphics[width=0.5\textwidth, angle=0]{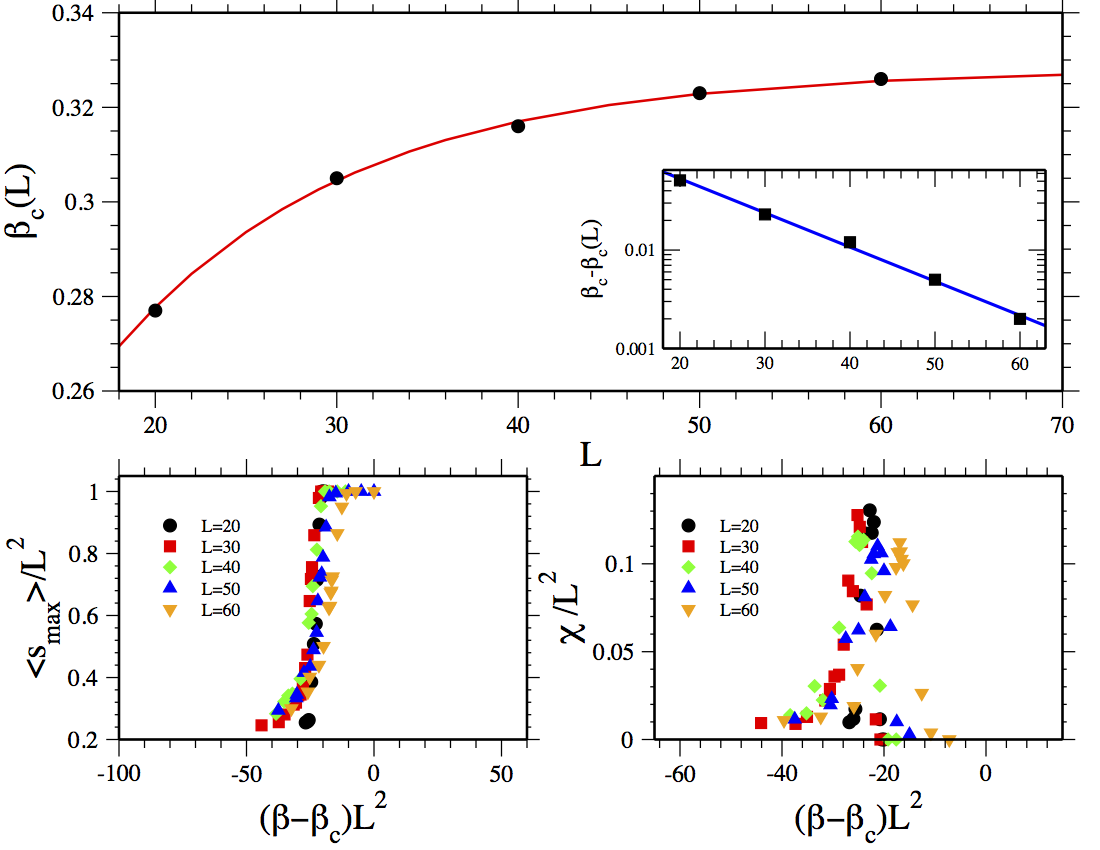}
\end{center}
\caption{\small 
\textit{Top}: Convergence to the asymptotic transition point $\beta_c$ 
of the finite-size transition points $\beta_c(L)$ measured from the
jump location in the normalized size of the largest cluster. 
The continuous line represents the best fitting function:
$\beta_c(L)=0.328-0.23 \exp(-0.076 L)$.
In the inset, the semilogaritmic plot shows in details the expected exponential behavior of the difference $\beta_c-\beta_c(L)$.
The continuous line represents an exponential fitting.
\textit{Bottom}: 
Rescaled plot for $s_{max}/L^2$ (on the left) and its fluctuations (on the right). 
}
\label{fig_rescaled}
\end{figure}

An alternative approach \cite{borgs} for the estimation of the 
asymptotic transition point uses 
the location of the observed discontinuity 
in  the normalized size of the largest cluster
(the first value of $\beta$ for which $<s_{max}>/L^2$
is less than $1$).
Using this method the convergence is supposed to be exponential
and the extrapolation for $L\to \infty$ gives $\beta_c= 0.328\pm0.002$ (see
Figure \ref{fig_rescaled}).  
In the same figure we can observe how the difference $\beta_c-\beta_c(L)$ 
clearly presents the expected exponential behavior. 

Additional consistency checks of the above results can be performed
verifying if  the measured quantities present  
the typical scaling of a discontinuous transition near the 
transition point, a standard procedure for equilibrium finite-size scaling analysis \cite{binder}.
The scaling plot of $<s_{max}>/L^2$     
should be obtained simply 
considering the rescaled control parameter $\beta^*=  (\beta-\beta_c)L^{d}$,
where $d$ is the system dimension.
In a similar fashion, the scaling plot of the fluctuations should be obtained
considering the rescaled fluctuation $\chi \cdot L^{-d}$ and the rescaled  parameter $\beta^*$.
As shown in details in Figure \ref{fig_rescaled},
it is possible to obtain a reasonable collapse which satisfies these relations.
In fact, data roughly collapse to a single curve, strongly suggesting the validity of the finite-size scaling ansatz expected for a discontinuous transition.

\begin{figure}[t]
\begin{center}
\includegraphics[width=0.5\textwidth, angle=0]{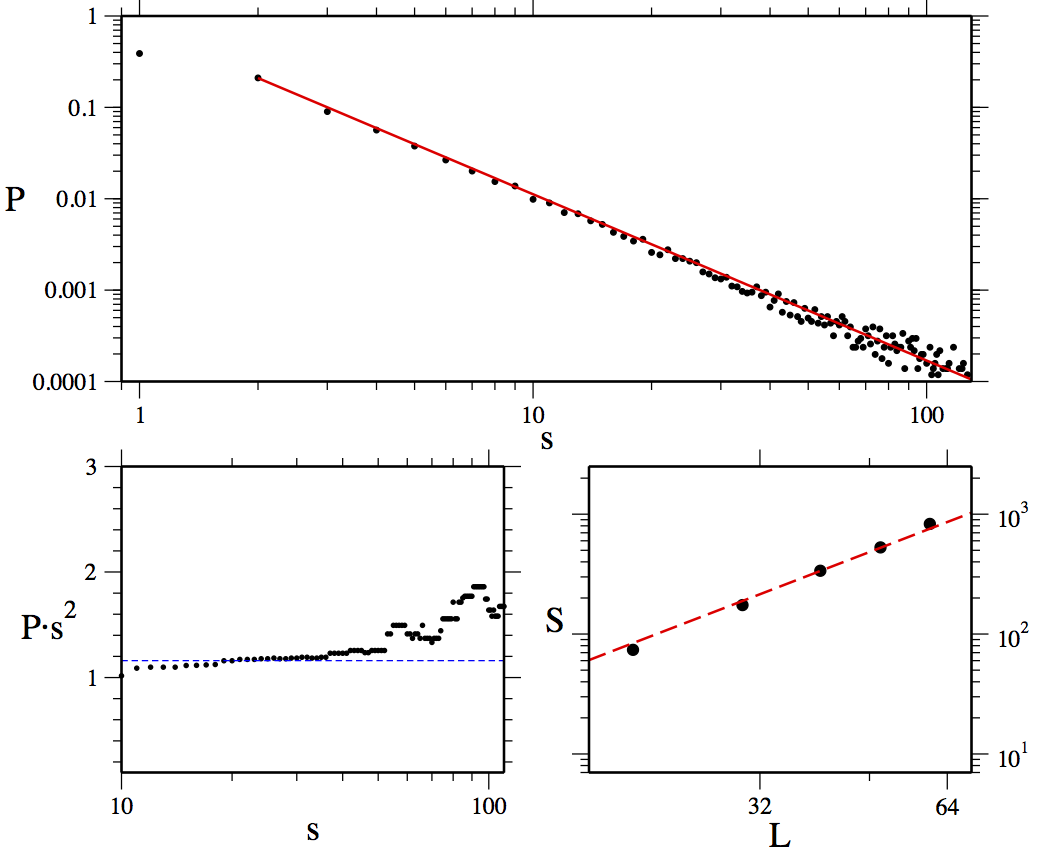}
\end{center}
\caption{\small 
{\it Top}: Probability distribution of the size of clusters $P(s)$
for 
$\beta=0.3235$ and $L=60$. 
The continuous line represents the power-law fitting of the data.
{\it Bottom}: On the left, $P(s)\cdot s^2 $ as a function of $s$.
The probability distribution is binned and it shows 
an exponent very close to $2$ at the intermediate region of $s$, just before the development of the finite size effects.
On the right, the average cluster size $S$ is plotted as a function of $L$ at $\beta_c(L)$. The dashed line has slope 2. Results have been averaged over 100 samples.
}
\label{fig_size}
\end{figure}

We conclude our study analyzing the behavior of the size distribution of the clusters present in the fragmented phase near the transition.
Obviously we examine only the simulations which do not converge to a unique cluster and, among them, the domains characterized by agents sharing the same unique convention 
(the regions with memory containing more than one convention 
are not considered).
The probability distribution of the clusters size $s$ 
is presented in Figure \ref{fig_size}.
In the range of the system size we study, the distribution decays as a power law. 

An accurate estimation of the exponent is difficult for the system size 
we are using. 
A data fitting allover the region of considered $s$ gives  an exponent  near $1.8$. 
However, the exponent is very close to $2$ in the intermediate region of $s$, just before the finite size effects which 
generate an accumulation on the higher bins responsible 
for the exponent decrease (see Figure \ref{fig_size}).
For an unambiguous characterization of this behavior
and for the identification of possible logarithmic
corrections to a consistent power law slope,
simulations on larger systems should be realized.
We remember that the fact of finding a discontinuous phase transition 
with a power law clusters size distribution, and not an exponential
or Gaussian one, is not an unusual result, as can be seen, 
for example, in the Axelrod's model for social influence \cite{castellano00}.

In analogy with the approach used in percolation theory, we can measure
the average cluster size $S$ defined as:
\begin{eqnarray}
S=\frac{\sum_{s}n_{s}s^2}{\sum_{s}n_{s}s}
\nonumber
\end{eqnarray}
where $n_s$ stands for the number of clusters of size $s$ and the sum run over all possible values of $s$. 
For a discontinuous transition $S$  is expected to scale with $L^2$ at the transition point \cite{Fortunato}.
This relation is reasonable verified for our system as can be stated looking at Figure \ref{fig_size}.\\


In summary, we have presented a numerical study of the 
Naming Game model with a noise term on a regular 2D lattice
with the intent of clearly characterize, for the first time,
the nature of the transition generated by this system.


Our analysis has shown that the model effectively displays a
non-equilibrium phase transition between
a consensus state and a phase with coexisting conventions in dependence
of the control parameter which represents the efficiency of the communication
process.
The nature of the phase transition has been studied by means of a
finite-size scaling analysis. 
The variance and the moment ratio show a scaling 
behavior which can be associated with
a  discontinuous transition  
and allow the estimation of the transition point at the 
thermodynamic limit.
Additional confirmations of these results come from 
the collapse of the scaling plot of $<s_{max}>/L^2$
and its fluctuations which have been obtained using the scaling 
law expected for a discontinuous transition.

Finally, we have presented some results for the behavior of the clusters distribution near 
the transition point. The distribution displays a power law behavior, while
the average cluster size scaling with the system size is compatible with a
discontinuous transition. \\

These results are not relevant only for the
naming game community, but, more in general, they can be
interesting for the  community of the statistical physicists which work with
  discontinuous nonequilibrium phase transitions into absorbing states.
In particular, they are related to  systems used in the description
  of opinion formation and which are characterized
  by the presence  of a large number of possible different absorbing states.
Considering systems embedded in a 2D space, we can mention 
the Axelrod’s model.
  In such a model, the consensus-fragmentation phase
  transition is discontinuous if the number of cultural features
  is greater than two \cite{castellano00}.
A discontinuous behavior was also described in another
  version of the Naming Game model, characterized by the presence
  of an open-ended reservoir of words \cite{brigatti15}.




\end{document}